\documentstyle[11pt,paspconf,epsf,psfig,twoside]{article}

\newcommand{\pmin}{P_{\rm min}}
\newcommand{\msun}{M_{\sun}}

\begin{document}

\title{The CV period minimum}


\author{Ulrich Kolb}
\affil{Astronomy Group, University of Leicester, Leicester LE1 7RH, U.K.}
\author{Isabelle Baraffe}
\affil{C.R.A.L.\ (UMR 5574 CNRS), Ecole 
Normale Sup\'{e}rieure de Lyon, F-69364 Lyon Cedex 0.7, France}

\begin{abstract}
Using improved, up--to--date stellar input physics tested against
observations of low--mass stars and brown dwarfs we calculate the
secular evolution of 
low--mass donor CVs, including those which form with a brown dwarf
donor star. Our models confirm the mismatch between the calculated
minimum period ($\pmin \simeq 70$~min) and the observed short--period
cut--off ($\simeq 
80$~min) in the CV period histogram. Theoretical period distributions
synthesized from our model sequences always show an accumulation of
systems at the minimum period, a feature 
absent in the observed distribution. We suggest that non--magnetic CVs
become unobservable as they are effectively trapped in permanent
quiescence before they reach $\pmin$, and that small--number
statistics may hide the period spike for magnetic CVs.  
We calculate the minimum period for high mass transfer rate sequences
and discuss the relevance of these for explaining the location of CV 
secondaries in the orbital period - spectral type diagram. We also
show that a recently suggested revised mass--radius relation for 
low--mass main--sequence stars cannot explain the CV period gap.
\end{abstract}


\section{Introduction}
We consider cataclysmic variables (CVs) at period--bounce, the
evolutionary phase where the secular mean orbital period derivative
changes from negative to positive. For predominantely convective donor
stars (with masses $\la 0.6-0.7 \msun$) period bounce occurs
when the mass transfer timescale $t_M = M_2/(-\dot M_2)$ becomes small
compared to the secondary's thermal time $t_{\rm KH}=GM_2^2/R_2L_2
\simeq 3 \times 10^7 {\rm yr}/(M_2/\msun)^3$. These stars expand and become
less dense on rapid mass loss, 
in contrast to the increase of mean
density $\rho=M_2/R_2^3$ towards smaller mass along the main
sequence. When $\rho$ decreases 
the orbital period becomes longer as $P \propto \rho^{-1/2}$ (from
Roche geometry and Kepler's law).

The short--period cut--off of the CV orbital period distribution at 
$80$~min, the ``minimum period'', has long been interpreted as a
consequence of period bounce (Paczy\'nski \& Sienkiewicz 1981;
Rappaport, Joss \& Webbink 1982) coinciding with the donor's
transition from a main--sequence star to a brown dwarf (BD). 
Period bounce occurs at longer orbital period and correspondingly 
higher donor mass if the mass transfer rate is much higher than
the rate driven by gravitational wave emission. 
CVs are close to period bounce at the upper edge of the CV
period gap for transfer rates required to fit the 
width and location of the gap within the standard period gap model.
Yet higher rates and corresponding bounce at still longer $P$ may
be required to explain the observed location of certain 
CVs in the orbital period - spectral type diagram (Beuermann et
al.\ 1998).

Here we investigate the behaviour of CVs at period bounce by applying
up--to--date stellar models for low--mass stars and brown dwarfs
(Baraffe et al.\ 1995, 1997, 1998; henceforth summarized as BCAH)
which reproduce observed properties of field M--dwarfs with
unprecedented accuracy. Main features of these models 
are the improved internal physics (Chabrier and Baraffe 1997) and
outer boundary conditions based on non-grey {\it NextGen} atmosphere
models and synthetic spectra of Hauschildt et al.\ (1998; see also
Allard et al.\ 1997).

\section{The minimum period at $80$~min}

\begin{figure}[t]
\psfig{file=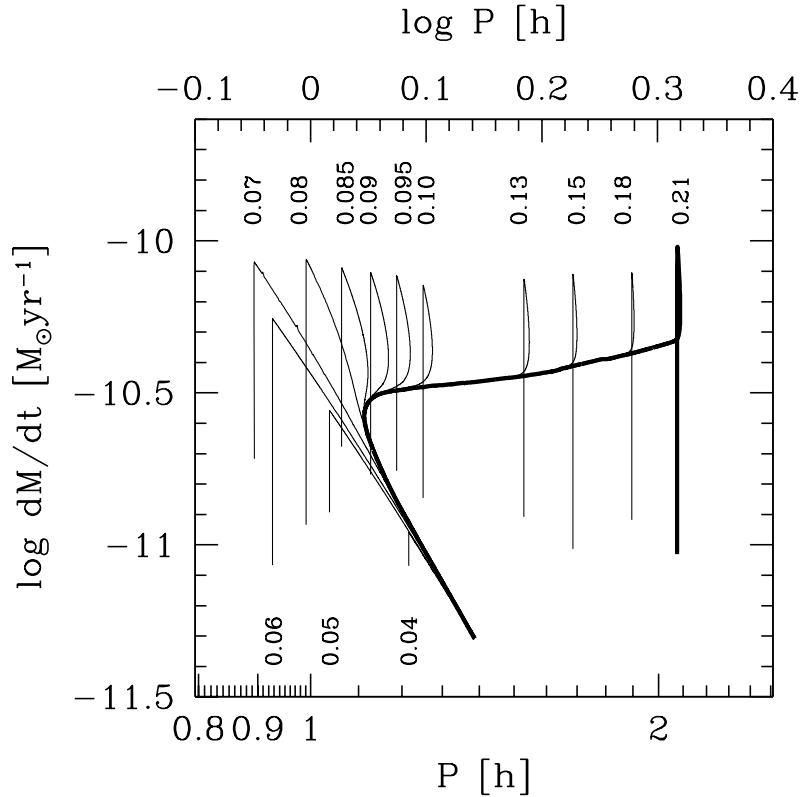,width=12cm}
\vspace{-1cm}
\caption{Mass transfer rate versus orbital period (left) and donor's
effective temperature versus period (right) along evolutionary 
sequences for CVs with a $0.6 \msun$ WD. The tracks are labelled with
the initial donor mass (in $\msun$). The $0.21\msun$ sequence is shown
in bold.
\label{kb1}}
\end{figure}

\begin{figure}[t]
\psfig{file=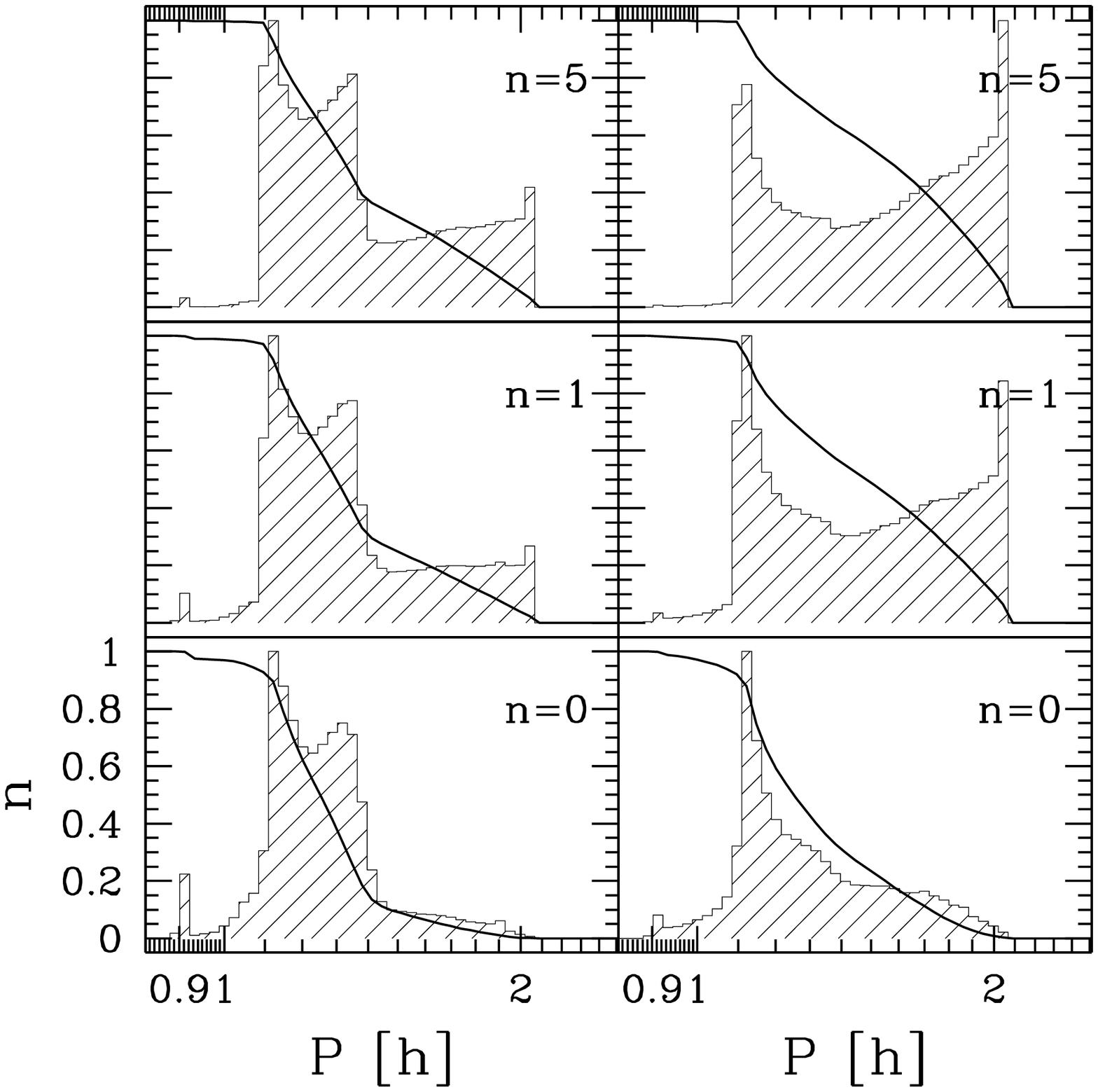,width=12cm}
\vspace{-1cm}
\caption{Theoretical orbital period distribution (histogram and
cumulative distribution) for CVs with $0.6 \msun$ WD mass and donor
mass $\le 0.21\msun$ (i.e.\ period $P<2.1$~h), for different
contributions $n$ from systems born 
above the period gap. Left: volume--limited sample ($\alpha=0$). Right:
magnitude--limited sample ($\alpha=1.5$). See text for details. 
\label{kb2}}
\end{figure}

The observed period distribution of hydrogen--rich donor CVs shows a
sharp cut--off at $P \simeq 80$~min (e.g.\ Ritter \& Kolb 1998), the
short end being marked by J0132--6554 at $P=77.8$~min. The notable
exception is V485 Cen at $P=59$~min, while six AM CVn type CVs with yet 
shorter periods are interpreted as helium--star CVs.

\subsection{Model calculations}
Theoretical period distributions calculated by population synthesis
techniques (e.g.\ Kolb 1993, 1996; Howell, Rappaport \& Politano 1997)
so far fail to reproduce the location of the minimum period $\pmin$
(they typically give a cut--off at $60 - 65$~min), and predict a
significant accumulation of systems 
near $\pmin$, a ``period spike'', which is not observed. The spike is
caused by the fact that the detection probability $d$ per period bin is
inversely proportional to the velocity $\dot P$ in period space, hence
$d \rightarrow \infty$ at period bounce $\dot P = 0$.

The population models rely on a number of simplifications. The donor
is usually approximated as a polytrope, with a surface
boundary condition calibrated to full stellar models.
In the case of Kolb (1993) these were obtained using Mazzitelli's
stellar code as of 1989 (e.g., Mazzitelli 1989), with now partly
out--dated stellar input physics.
Equally important, CVs forming with donor mass $M_2 \la 0.1
\msun$ are not considered. All CVs with smaller $M_2$ in the
population models descended from systems with initially 
higher--mass donors.

To improve on this we recalculated the secular evolution of
short--period CVs using the BCAH stellar evolution code. We focussed on
CVs below the period gap and assumed that orbital angular momentum is
lost by gravitational wave emission (e.g.\
Landau \& Lifschitz 1952) and via an isotropic stellar wind from the
white dwarf (WD), removing the accreted mass with the WD's specific
orbital angular momentum from the binary. With this choice the  
results do not depend on the rather uncertain strength of magnetic 
braking thought to dominate the evolution of CVs above the period gap
(see e.g.\ King 1988 for a review). 
 
In Fig.~\ref{kb1} we plot mass transfer rate versus period for
a set of sequences with $0.6\msun$ WD mass and initial donor
masses ranging from $M_2 = 0.04\msun$ to $0.21\msun$. Within the
standard period gap model (e.g.\ Kolb 1996) the upper value
is the donor mass at which CVs emerge from the detached
phase that is responsible for the CV period gap. 
At turn--on of mass transfer the secondary was
either on the ZAMS ($M_2>0.09\msun$), or had an age of 2 Gyrs
($M_2\le 0.09\msun$).  

The figure confirms the well known effect that systems with different
initial donor mass join rather quickly a uniform evolutionary track
(Stehle et al.\ 1996). Most systems undergo period bounce at $\pmin
\simeq 67$ min, which is only slightly longer than the corresponding
$\pmin=65$~min found with Mazzitelli's models (Kolb \& Ritter 1992). 
Significantly, CVs forming with fairly old and massive brown dwarf
donors (age $\ga 2$~Gyr, mass $0.05-0.07\msun$) would populate the
period regime shortwards of $\pmin$. V485 Cen could be such a
CV.

The importance of CVs forming with BD donors for the observed period
histogram depends on their relative formation rate. 
Here we consider the simple case of a constant birth rate $b$ per
logarithmic mass interval ($\partial b/\partial \log
M_2 = 0$) for $0.04 \le M_2/\msun \le 0.21$. In the case of CVs with
main--sequence secondaries this choice is 
roughly consistent with results from detailed calculations
of CV formation via the standard common envelope channel (e.g.\
Politano 1996, de~Kool 1992), while it is just an assumption for
degenrate secondaries.    
(A more detailed appraisal of the BD CV formation rate is in
preparation.) CVs which have formed above the
period gap and evolved to short periods are taken into account by
adding the contribution $n\times I$ to the birth rate in the first
mass bin at $0.21\msun$. This corresponds to a period $P\simeq2.1$~h,
so that the predicted distribution only applies to periods less than
this. Here $I$ is the total formation rate of
CVs below the period gap and $n$ a free parameter. Detailed  
standard models indicate $n\simeq1$. 

The period distribution for the subset of CVs with $0.6\msun$ WD mass
obtained from a convolution of the sequences 
in Fig.~\ref{kb1} with the above simple CV formation rate is shown in 
Fig.~\ref{kb2}, for $n=0$, $1$ and $5$. The left panel shows the
distribution of a volume--limited sample, the right panel a sample
where individual systems have been weighted by $\dot M^\alpha$, with
$\alpha=1.5$ ($\dot M$ is the mass transfer rate). With this
brightness--dependent factor the period distributions mimic those
expected for $m_{\rm vis}$-- or $m_{\rm bol}$--limited samples. The
proper value for the parameter $\alpha$ is of order unity, but
depends in detail on the distribution of objects in physical space and
the emission properties of accretion discs with accretion rate $\dot
M$ (cf.\ Kolb 1996). The population shown in the figure formally
corresponds to an age of 6~Gyr for the Galactic disc; in a somewhat 
older population the edge of the volume--limited distributions at
$\simeq 1.4$~hr would appear at slightly longer $P$.   

Figure~\ref{kb2} shows that for the adopted formation rate BD CVs
would not contribute significantly to the period distribution in a
magnitude--limited sample. In particular, BD CVs have no effect on the
overall shape of the period spike at $\pmin$. They would have an
effect only if $b$ were strongly increasing towards small
masses. In this case the short--period end of the
distribution would be at even shorter periods, clearly in conflict 
with observations. 

The overall shape of the weighted distribution is not sensitive to
$\alpha$; there is hardly any difference between   
the cases with $\alpha=1$ and $\alpha=1.5$. Increasing $\alpha$ tends
to decrease the amplitude of the spike, but we obtain a distribution
which is roughly flat -- similar to the observed one -- only for
$\alpha \ga 6$. This holds also for calculated period distributions
which take into account the full WD mass spectrum expected for CVs
(see Kolb \& Baraffe 1998 for an example obtained with deKool's (1992)
CV formation rate).  

In summary: suppressing the period spike in the distribution
requires a very steep dependence of the detectability on $\dot M$. 

\subsection{Interpretation}

Essentially all known short--period non--magnetic CVs are dwarf novae,
with similar absolute magnitudes in outburst (e.g.\ Warner
1995). A few of them, sometimes referred to as WZ Sge stars,
have very long outburst recurrence times $t_{\rm rec}$, the most
extreme example being WZ Sge itself with $t_{\rm rec}\simeq30$~yr.
It has long been noted that low--$\dot M$ CVs might have escaped
detection if their outburst interval is significantly longer than
the period since beginning of systematic monitoring and surveying
of the sky with modern means -- a few decades.

For long $t_{\rm rec}$ the relative detection probability of
dwarf novae scales as $d \propto 1 / t_{\rm rec}$,
suggesting that $t_{\rm rec} \propto \dot M^{-\alpha} \propto \dot
M^{-6}$ or steeper. 
In practice this 
very steep dependence would require that $t_{\rm rec} \rightarrow
\infty$ for $\dot M \la {\rm few} \times 10^{-11} \msun$yr$^{-1}$, i.e.\
low--$\dot M$ CVs would not undergo outbursts at all.  

A plausible physical model which naturally accounts for this property
is the extreme version of the evaporating accretion disc model
suggested for WZ~Sge  
by Meyer--Hofmeister et al.\ (1998; see also Liu et al.\ 1997). 
Evaporation of accreted material into a hot corona could prevent the
disc from accumulating the critical surface mass density required to
launch a heating wave. Systems in this permanent quiescence are
optically very faint but should emit about $10\%$ of their accretion
luminosity $L_{\rm acc} = G M_1 \dot M / R_1$ ($M_1, R_1$ is the
WD's mass and radius) in X--rays. 
See Watson (this volume) for a plot of
the corresponding intrinsic X--ray luminosity function of a typical
Galactic CV population (model dK1in of Kolb 1993), 
and for a discussion of 
the detectability of such an X--ray background from low--luminosity
CVs and its potential contribution to the Galactic ridge emission.

\begin{figure}
\begin{minipage}{55mm}
\psfig{file=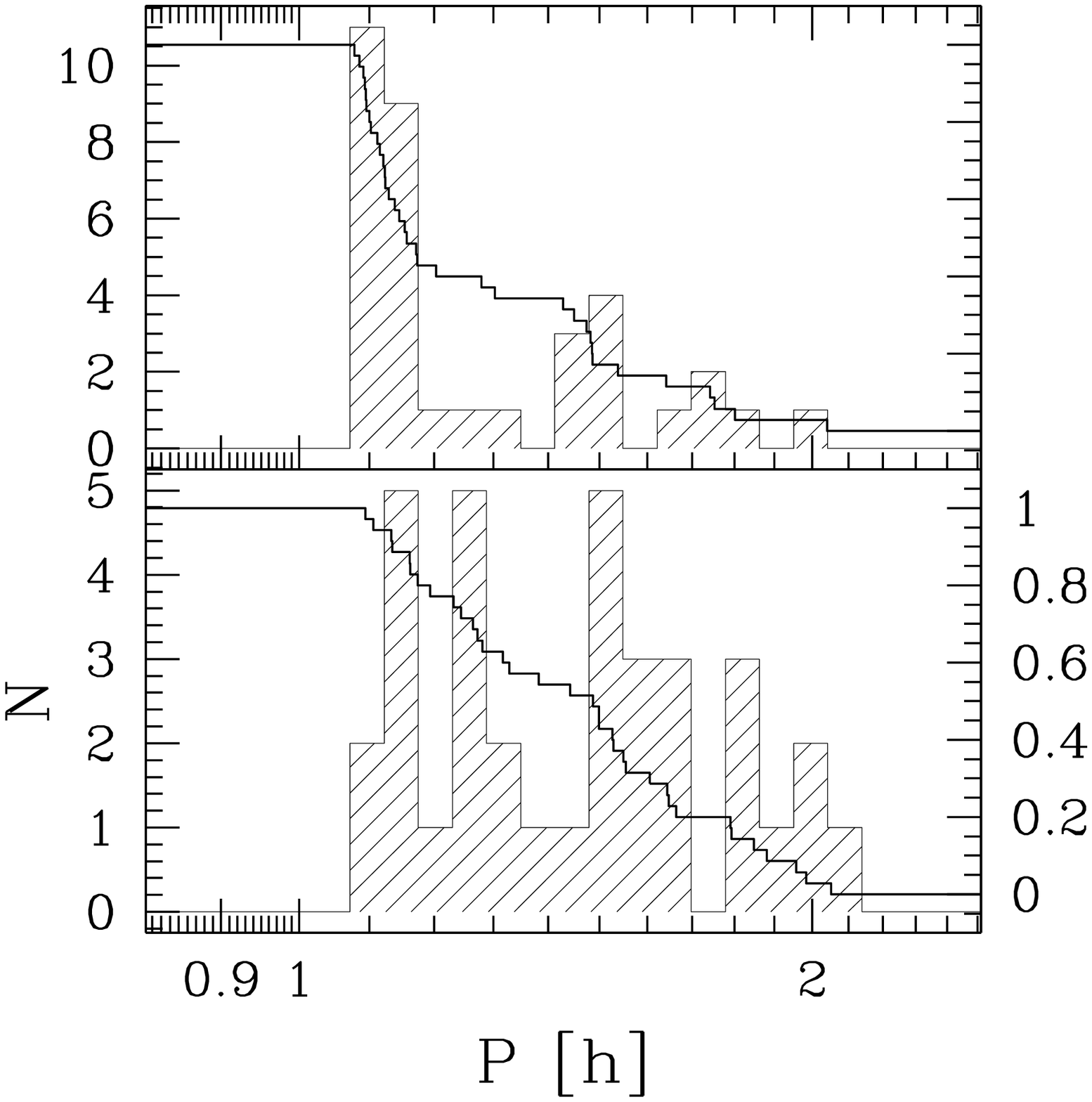,width=55mm}
\end{minipage}
\hspace*{1cm}
\begin{minipage}{55mm}
\psfig{file=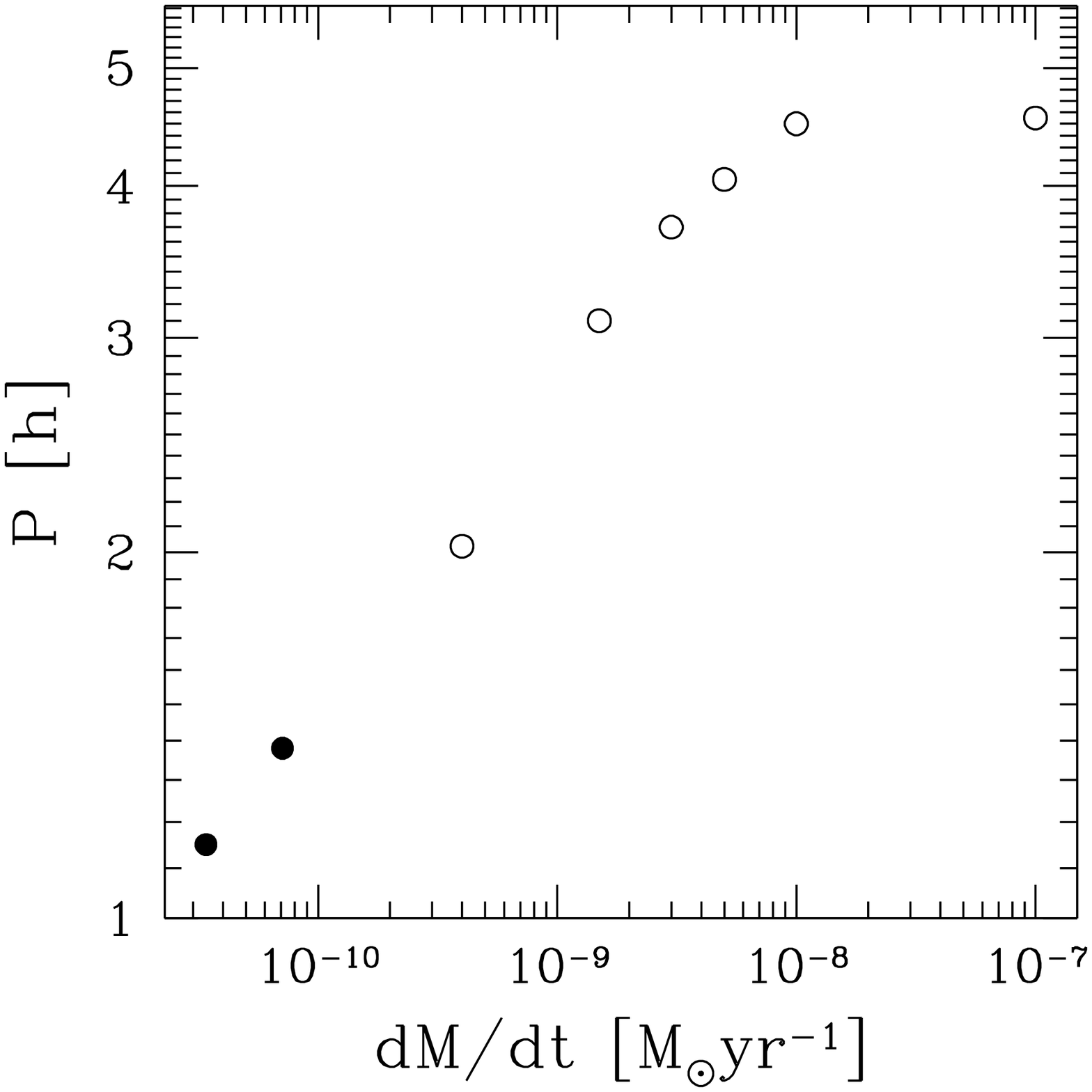,width=55mm}
\end{minipage}
\caption{{\em Left:} Two representative samples of 35 systems drawn 
from the parent distribution shown in Fig.~\protect\ref{kb2}, middle
right. 
{\em Right:} Orbital period at period bounce versus mass transfer rate
at this point. Solid dots represent sequences calculated with angular 
momentum loss $\dot J=\dot J_{\rm GR}$ and $4\times \dot J_{\rm GR}$,
open dots are sequences calculated with a constant mass transfer
rate. 
\label{kb4}}
\end{figure}

Although this could explain the non--detection of the period spike
at $\pmin$ for dwarf novae, it certainly would not apply to discless
systems, i.e.\ for the $\simeq 35$ polars with $P\la2$~h (Beuermann
1997; Ritter \& Kolb 1998). An independent observational selection
effect operating on 
polars in the same period/mass transfer rate range and with the same
net result as a steep increase of $t_{\rm rec}$ for dwarf novae
seems highly unlikely. 
In other words: polars should show a period spike, even though dwarf
novae do not. Small number statistics could be responsible for the
non--detection of the polar spike. Monte Carlo experiments where a
sample of 35 systems is drawn from an underlying period distribution
like the one shown in 
Fig.~\ref{kb2} (middle right) give a surprisingly wide variety of
distributions (Fig.~\ref{kb4}, left), with many of them showing no sign of a
period spike at all. In contrast, in a sample with $\ga 80$ systems
(dwarf novae with $P<2.1$h) the spike is almost always prominent.  

This explanation for the missing period spike allows the theoretically
computed value of $\pmin$ to be shorter than the observed one.
Nelson et al.\ (1985) found 
that their low--mass stellar models increase the computed minimum
period by $\simeq 10\%$ when tidal and rotational corrections are
applied to the 
1--dim.\ stellar structure equations (Chan \& Chau 1979). 
Preliminary calculations based on the BCAH models 
do not confirm this result; we find a rather modest increase of
$\simeq 1\%$ (cf.\ Kolb and Baraffe 
1998 for details).
While we do not claim that the true theoretical value of $\pmin$ must
be longer than what our models give, there are indeed uncertainties in
the calculated value of $\pmin$ inherent to  
the very concept of the Roche model. Strictly valid only for point
masses, its applicability to extended stars relies on the fact 
that stars are usually sufficiently centrally condensed. This is not
necessarily a good approximation for fully convective stars which are
essentially polytropes of index $3/2$.
 
Finally, we note that the predicted $\pmin$ is longer if the orbital 
angular momentum losses $\dot J$ are larger than the value 
$\dot J_{\rm GR}$ from gravitational radiation.
We find $\pmin \simeq 83$~min (up from $69$~min) for $\dot J = 4 \times
\dot J_{\rm GR}$ and $M_1=1\msun$, a much smaller increase than quoted
by Paczy\'nski (1981).  
Patterson (1998) favoured a modest increase of $\dot J$ over $\dot
J_{\rm GR}$ on grounds of space density considerations and 
the number ratio of CVs above/below the gap. However, postulating an
as yet unknown $\dot J$ mechanism which conspires to produce almost the 
same value as $\dot J_{\rm GR}$ at the transition from non--degenerate to
degenerate stars does not seem very attractive. 

A more thorough account of the discussions in this section is given in
Kolb \& Baraffe (1998).

\section{Period bounce at longer periods}

\begin{figure}[t]
\psfig{file=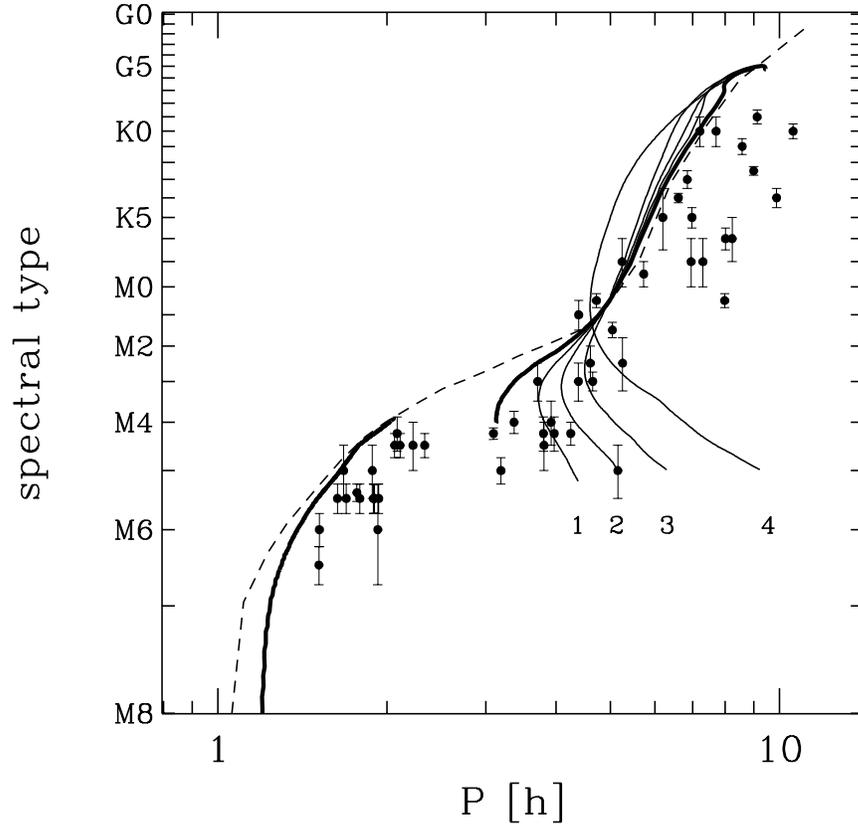,width=12cm}
\vspace{-1cm}
\caption{Spectral type of the secondary versus orbital period. Data
points from Beuermann et al.\ 1998; dashed: ZAMS; heavy solid line:
standard sequence (see text); solid: sequences with constant mass
loss rate (1: $3 \times 10^{-9} \msun$yr$^{-1}$; 2: $5 \times 10^{-9}
\msun$yr$^{-1}$; 3: $10^{-8} \msun$yr$^{-1}$; 4: $10^{-7}
\msun$yr$^{-1}$).  
\label{kb5}}
\end{figure}

Reversal of the orbital period derivative is not uniquely confined to
the transition region between main--sequence and degenerate
stars. Period bounce occurs at longer periods and higher donor masses
if the system has a higher mass transfer rate than the one driven
by gravitational radiation. As an example, we show in Fig.~\ref{kb4}
(right) 
how $\pmin$ varies with the transfer rate.

The potential importance of evolutionary sequences with fairly high
mass transfer rates become clear from Fig.~\ref{kb5} (right) where we
plot the donor's spectral type versus orbital period (data taken from
Beuermann et al.\ 1998) and  a ``standard'' evolutionary sequence that
reproduces the width and location of the period gap ($2.1-3.2$~h).
This standard sequence was calculated with 
constant mass transfer rate $1.5 \times 10^{-9} \msun$yr$^{-1}$
until the secondary became fully convective, when mass loss
was terminated and the star allowed to shrink back to its equilibrium
radius. Mass loss resumed (at the now shorter orbital period
$P=2.1$~hr) with a rate $5 \times 10^{-11} \msun$ yr$^{-1}$, typical
for mass transfer driven by gravitational wave emission.

Although the BCAH models successfully reproduce properties of
low--mass single stars (Baraffe \& Chabrier 1996, Baraffe et al.\
1997, 1998, and references 
therein) the standard sequence clearly fails to give the observed late
spectral types for a given period above the gap.
As has been shown
by Beuermann et al.\ (1998) CVs with donors that are somewhat evolved
off the ZAMS (but still in the core hydrogen--burning phase) can account
for the late spectral types in long--period systems
($P\ga 6$~h), but not in those close to the upper edge of the gap
($3\la P/{\rm h}\la 6$) where evolved and unevolved sequences merge.
(Further details will be presented in Baraffe \& Kolb 1998). 

Generally, low--mass stars in the phase of core hydrogen burning 
subjected to mass loss become over- or underluminous compared to
stars in thermal equilibrium, but in such a way that the effective
temperature hardly changes (King \& Kolb 1998). Hence the
effective temperature, or the spectral type, is an indicator of
the stellar mass, whatever the previous mass--loss history. 

If the observed late spectral types of CV secondaries for $P\la 6$~h
are a consequence of mass transfer alone, then these secondaries
must be severely undermassive, i.e.\ less massive than a 
main--sequence star that would fill its Roche lobe at the same 
period. As the degree of ``undermassiveness'' increases with mass
transfer rate $\dot M$, we conclude that 
in systems with $3\la P/{\rm h}\la 6$ and cool donors,
$\dot M$ must be higher than in the standard sequence.
As an extreme example, AR Cnc with spectral type M5
and $P=5.15$~h would fit on a track with $\dot M = 5 \times
10^{-9} \msun$yr$^{-1}$, at a donor mass $\simeq 0.15 \msun$. (We note
that this transfer rate is only marginally consistent with the fact
that AR Cnc is a dwarf nova). With such a small donor mass and high
transfer rate this system does not fit 
into the framework of the standard model of the period gap, where  
the orbital decay rate drops sharply at $M_2\simeq0.2\msun$. If a
significant fraction of systems close to the 
upper edge of the gap were similar to AR Cnc, the period gap would not
show up in the collective period distribution. 
To verify if this is indeed a problem for the standard CV period gap
model, an observational determination of the donor mass in systems
like AC Cnc is vital.

\section{Main--sequence mass--radius relation and CV period gap}

As highlighted in the previous section the spectral types of certain CV
secondaries, when interpreted as a result of their mass loss history,
may be in conflict with the standard CV period gap
model. Numerous alternative explanations for the period gap have been
put forward in the literature, but none of 
them has proved to be as successful as the disrupted orbital decay model.
The most recent alternative suggestion, by Clemens et al.\ (1998), 
claims that a characteristic feature of the low--mass main--sequence
mass--radius relation $R(M)$ would translate into a rapid evolution
through the period gap region even for continuous angular momentum
losses, 
reducing  the discovery probability there. In particular, the
logarithmic slope $\zeta = {\rm d}\ln R/{\rm d}\ln M$ of
Clemens et al.'s observationally derived $R(M)$ relation shows 
two discontinuities. With decreasing mass $\zeta$ jumps from a rather
large value $\ga 1$ to a rather small $\zeta\simeq 0.33$ at mass
$\simeq 0.5\msun$ and $\simeq 0.2\msun$. A corresponding 
structure may already be evident in the underlying colour--magnitude
diagram, with disputable significance.
Note that theoretical models, which now reach a good
agreement with observed stellar parameters, e.g.\ the eclipsing binary
CM Draconis (Chabrier and Baraffe 1995) or the data of
Leggett et al.\ 1996, do not predict such a feature in the $R(M)$
relationship of low-mass stars. 

If mass transfer in CVs proceeds sufficiently slowly (as assumed by
Clemens et al.\ 1998), the donor
star's radius simply follows the main--sequence $R(M)$ relation. 
As 
\begin{equation}
\frac{\dot P}{P} = \frac{3}{2} \frac{\dot R_2}{R_2} - \frac{1}{3} 
                   \frac{\dot M_2}{M_2} = \frac{1}{2} \left( 3\zeta -
                   1 \right) \frac{\dot M_2}{M_2} ,
\end{equation}
the discovery probabilty $d (\log P) \propto  \dot M/(\dot P/P)$ in a 
magnitude--limited sample is $d \propto 1/(3\zeta - 1)$.   
Hence the $R(M)$ relation claimed by Clemens et al.\ would 
produce a pair of period ``spikes'' at 3.4~hr and 2.0~hr where $\zeta
\simeq 1/3$ ($d\rightarrow \infty$). (The CVs are actually close to
bounce at these periods, and the spikes appear for the same reason as
the spike at $\pmin$ discussed in Sect.~1.) 
The discovery probability between the two spikes is no lower than
outside them, there is 
no period ``gap''. This has been verified with a full population
synthesis calculation, adopting continuous, weak orbital
decay by gravitational radiation only, with the claimed $R(M)$
relation imposed on lower main--sequence stars (Kolb et al.\ 1998).

\acknowledgments
Theoretical astrophysics research at Leicester is supported by a PPARC
Rolling Grant. We thank Klaus Beuermann, Emmi Meyer--Hofmeister,
Friedrich Meyer, Mike Politano, Hans Ritter, Rudi Stehle, Mike Watson
and Peter Wheatley  for discussions, and Andrew King for a critical
reading of the manuscript. UK acknowledges travel support from
NASA/GSFC.

\end{document}